\documentstyle[aps,pra,eqsecnum,psfig,multicol,graphicx]{revtex}

\begin{document}
\draft
\title{Weak Values, Quantum Trajectories, and 
the Stony-Brook Cavity QED experiment}
\author{H.M. Wiseman}
\address{Centre for Quantum Dynamics, 
School of Science, Griffith University, Brisbane, Queensland
4111, Australia}
\date{\today}
\maketitle

\begin{abstract}
Weak values as introduced by Aharonov, Albert and Vaidman (AAV)
are ensemble average values for the results of weak 
measurements. They are interesting when the ensemble is preselected on 
a particular initial state and postselected on a particular final 
measurement result. I show that weak values  arise naturally in quantum 
optics, as weak measurements occur whenever an open system is 
monitored (as by a photodetector). I use quantum trajectory theory to 
derive a generalization of AAV's formula to include 
 (a) mixed initial conditions, (b) nonunitary evolution,  (c) a 
 generalized (non-projective) final measurement, and (d) a 
 non-back-action-evading weak measurement. I apply this theory to the 
 recent Stony-Brook cavity QED experiment demonstrating wave-particle 
 duality [G.T. Foster, L.A. Orozco, H.M. Castro-Beltran, and H.J. Carmichael,
Phys. Rev. Lett. {\bf 85}, 3149 (2000)]. 
I show that the ``fractional'' correlation function measured in 
 that experiment can be recast as a weak value  in a form as simple as 
that introduced by AAV.
\end{abstract}


\newcommand{\beq}{\begin{equation}}
\newcommand{\eeq}{\end{equation}}
\newcommand{\bqa}{\begin{eqnarray}}
\newcommand{\eqa}{\end{eqnarray}}
\newcommand{\nn}{\nonumber}
\newcommand{\nl}[1]{\nn \\ && {#1}\,}
\newcommand{\erf}[1]{Eq.~(\ref{#1})}
\newcommand{\rf}[1]{(\ref{#1})}
\newcommand{\dg}{^\dagger}
\newcommand{\rt}[1]{\sqrt{#1}\,}
\newcommand{\smallfrac}[2]{\mbox{$\frac{#1}{#2}$}}
\newcommand{\half}{\smallfrac{1}{2}}
\newcommand{\bra}[1]{\langle{#1}|}
\newcommand{\ket}[1]{|{#1}\rangle}
\newcommand{\ip}[2]{\langle{#1}|{#2}\rangle}
\newcommand{\schx}{Schr\"odinger }
\newcommand{\sch}{Schr\"odinger}
\newcommand{\heix}{Heisenberg }
\newcommand{\hei}{Heisenberg}
\newcommand{\bl}{{\bigl(}}
\newcommand{\br}{{\bigr)}}
\newcommand{\ito}{It\^o }
\newcommand{\str}{Stratonovich }
\newcommand{\dbd}[1]{\frac{\partial}{\partial {#1}}}
\newcommand{\sq}[1]{\left[ {#1} \right]}
\newcommand{\cu}[1]{\left\{ {#1} \right\}}
\newcommand{\ro}[1]{\left( {#1} \right)}
\newcommand{\an}[1]{\left\langle{#1}\right\rangle}
\newcommand{\st}[1]{\left| {#1} \right|}
\newcommand{\implies}{\Longrightarrow}
\newcommand{\bfi}{{\bf I}_{[0,t)}}

\begin{multicols}{2}

\section{Introduction}

The concept of weak values in quantum mechanics was formulated by 
Aharonov, Albert and Vaidman (AAV) \cite{AhaAlbVai88}, using the earlier 
two-wave-function 
formalism of Ref.~\cite{AhaBerLeb64}. A weak value is the ensemble 
average value of a weak measurement result. A weak measurement is one that 
minimally disrupts the system, while consequently yielding a minimal 
amount of information about the observable measured. 
For a given initial system state, the 
ensemble average of weak measurement results is the same as 
for strong (i.e. projective) measurement results. Where weak measurements are 
interesting is when a {\em final} as well as an initial state is 
specified. Here the final state is the result of a second 
measurement (a strong one), so that the ensemble average is taken over 
a {\em post\-selected} ensemble, in which the desired result for the final 
measurement was obtained. 

Let the desired initial system state be $\ket{\psi}$, and let 
$\psi$ denote the event that this is successfully prepared. 
Let the desired final system state be $\ket{\phi}$, and let $\phi$ 
denote the event that the appropriate result 
 is obtained by the final projective measurement. 
Let the observable to be measured at an 
intermediate time have operator $\hat{X}$, and denote the result 
of the measurement $X_{n}$. Here the nature $n$ of the measurement 
could be strong ($n=s$) or weak ($n=w$). In the former 
case, which corresponds to a projective measurement, 
$X_{s}$ reproduces the statistics of $\hat{X}$. In the latter case, 
 $X_{w}$ does not, and indeed is not 
confined to the spectrum of $\hat{X}$. We do require however that 
in all cases $X_{n}$ be an unbiased estimator, in the sense that
\beq
\an{X_{n}}_{\psi} \equiv {\rm E}[X_{n}|\psi] = \bra{\psi}\hat{X}\ket{\psi}.
\eeq
Here E denotes expectation value, and the vertical line denotes ``given''.
The postselected value we desire is
\bqa
\hspace{-5ex}{\phantom{\an{X_{w}}}}_{\phi}\!\an{X_{n}}_{\psi} &\equiv& 
{\rm E}[X_{n}|\psi,\phi] = \sum_{x} x {\rm 
Pr}[X_{n}=x|\phi,\psi] \\
&=& \frac{\sum_{x} x {\rm Pr}[\phi|X_{n}=x,\psi]{\rm Pr}[X_{n}=x|\psi]}
{\sum_{y} {\rm Pr}[\phi,X_{n}=y|\psi]}. \label{genform}
\eqa
Here ${\rm Pr}[H]$ denotes the probability for event $H$.

For simplicity, assume for the moment that there is no evolution 
between measurements. Then for a strong intermediate measurement, the 
postselected average value of $X_{s}$ is 
\bqa
\hspace{-5ex}{\phantom{\an{X_{w}}}}_{\phi}\!\an{X_{s}}_{\psi} 
&=& \frac{\sum_{x} x {\rm Pr}[\phi|X_{s}=x]{\rm Pr}[X_{s}=x|\psi]}
{\sum_{y} {\rm Pr}[\phi|X_{s}=y]{\rm Pr}[X_{s}=y|\psi]} \\
&=& \frac{\sum_{x} \st{\ip{\phi}{x}}^{2} x\st{\ip{x}{\psi}}^{2}}
{\sum_{y}  \st{\ip{\phi}{y}}^{2}  \st{\ip{y}{\psi}}^{2}} .
\label{strongval}
\eqa
Here $\ket{x}$ denotes an eigenstate of $\hat{X}$.
In the final result there is no direct connection between the 
initial state and the final state. The intermediate strong 
measurement has destroyed that connection, as it enables the dropping 
of the conditional $|\psi$ from ${\rm Pr}[\phi|X_{s}=x,\psi]$ in 
\erf{genform}.
Basically there are no surprises here.

Contrast this result with that from a  weak measurement, as derived by 
AAV:
\beq \label{weakval}
\hspace{-5ex}{\phantom{\an{X_{w}}}}_{\phi}\!\an{X_{w}}_{\psi} = {\rm 
Re}\frac{\bra{\phi}\hat{X}\ket{\psi}}{\ip{\phi}{\psi}}.
\eeq
Aharonov and Vaidman derive this in  detail in Ref.~\cite{AhaVai90}, and it 
is also derived here, in Sec.~II, from a somewhat different perspective.
This formula appears to have a much more of a mysterious 
``quantum'' nature than 
\erf{strongval}, as it involves the initial and final states linearly 
(rather than bilinearly) on the numerator and denominator. As a 
consequence, weak values can have bizarre properties, especially when 
the outcome $\phi$ used for postselection is improbable 
\cite{AhaVai90}. For example, 
the weak value $\hspace{-5ex}{\phantom{\an{X_{w}}}}_{\phi}\!\an{X_{w}}_{\psi}$ 
may lie {\em outside} the range of 
eigenvalues for $\hat{X}$ \cite{AhaAlbVai88,AhaVai90}. Counterfactual paradoxes 
\cite{Har92} can also be expressed and resolved in this language \cite{Aha01,Mol01}.

It might be thought that weak values would be mainly of theoretical 
interest, because arbitrarily weak measurements would not arise 
naturally in experiments \cite{fn1}. 
But in fact arbitrarily weak measurements arise 
all the time, whenever a system is monitored (i.e. measured continuously 
in time). The 
measurement theory for monitored systems was developed by 
mathematical physicists \cite{Bel88,BelSta92,Bar90,Bar93}. It was subsequently 
rediscovered by workers in 
quantum optics \cite{DalCasMol92,GarParZol92,Car93} when 
it was found necessary to describe 
the conditioning of systems on an observed photocurrent. 
In quantum optics, the theory of 
continuous monitoring, especially for homodyne and heterodyne 
measurement, has come to be known as {\em quantum 
trajectory theory} \cite{Car93,WisMil93c}. 

In this paper I show in Sec.~II how the AAV  formula can 
be derived using the approach of quantum trajectory theory. 
I derive a simple generalization arising from removing the restriction 
that the weak measurement of $X$ be a quantum non-demolition 
measurement. In Sec.~III I derive a more 
complete generalization which allows for 
(a) mixed initial conditions; (b) non-unitary evolution 
between measurements; and (c) a non-projective final measurement. 
Within this context, the AAV formula can be seen to be a 
special case of the correlation functions that have been used in 
quantum optics for a long time \cite{WalMil94}. 
Finally, in Sec.~IV I show 
that the correlation function measured in the recent cavity-QED 
experiment at Stony-Brook \cite{FosOroCasCar00} can be reformulated 
as a weak value almost exactly as originally 
proposed by AAV. 

\section{Weak Values: the Special Theory}

Monitoring a quantum system  can be treated by considering a sequence  
of discrete weak measurements, each taking time $\delta t$, and letting 
$\delta t \to dt$. To obtain a sensible limit, the strength of each 
measurement must scale as $\delta t$. By this I mean that the 
post-measurement system state should, on average, be 
different from the pre-measurement state by an amount of order 
$\delta t$. 

To describe weak measurements, we require generalized 
(non-projective) quantum measurement theory 
\cite{Kra83,BraKha92,GarZol00}. Let the measurement result be the random 
variable $X_{n}$ as above. In the case of efficient 
measurements, the measurement can be described in terms of a set of {\em 
measurement operators} $\{\hat{M}_{x}\}_{x}$, each associated with a 
result $X_{n}=x$. The post-measurement conditioned state is given by
\beq
\rho_{x}(t+\delta t) = \frac{\hat{M}_{x}\rho(t)\hat{M}_{x}\dg }{{\rm 
Pr}[X_{n}=x]},
\eeq
where the probability for the result is
\beq
{\rm 
Pr}[X_{n}=x] = {\rm Tr}
[\hat{M}_{x}\rho(t)\hat{M}_{x}\dg].
\eeq
 By conservation of probability, the measurement operators 
must obey a completeness relation
\beq \label{complet}
\sum_{x} \hat{E}_{x} = \hat{1},
\eeq
where $\hat{E}_{x} = \hat{M}\dg_{x}\hat{M}_{x}$ is known as the {\em probability 
operator} or {\em effect} for the result $x$.

As above, we require the measurement of $\hat{X}$ to be unbiased, so that 
 $\an{X_{n}(t)}_{\rho(t)} = {\rm Tr}[\rho(t)\hat{X}]$. This is 
 satisfied by the following   
measurement operator, which also has a suitably weak effect on the system 
\cite{WisDio01}
\beq
\hat{M}_{x} = (2\pi / \delta t)^{-1/4}\exp(-x^{2}\delta t/2)
\sq{1 + \delta t(x\hat{c} - \hat{c}\dg \hat{c}/2 )}.
\eeq
Here $\hat{c}$ is an operator such that
\beq
\hat{c} + \hat{c}\dg = \hat{X}.
\eeq
For 
convenience we are measuring time in units such that $\hat{c}\dg 
\hat{c}\, \delta t$ is dimensionless. Note that by allowing $\hat{c}$ 
to be non-Hermitian we are allowing for non-back-action-evading measurements of 
$\hat{X}$, even ignoring any evolution between measurements.

The measurement operators satisfy a continuous version of the 
completeness relation 
(\ref{complet}) to order $(\delta t)^{2}$:
\beq
\int dx \hat{M}_{x}\dg \hat{M}_{x} = \hat 1 + O(\delta t^{2}),
\eeq
In the limit $
\delta t \to dt$, the correction vanishes and 
it is not difficult to show \cite{WisDio01} that 
the measurement result $X_{w}$ has statistics given by
\beq
X_{w}(t) = {\rm Tr}[\hat{X}\rho(t)] + \xi(t) .\label{here}
\eeq
Here is a Gaussian white noise term, defined by 
$\xi(t)=dW(t)/dt$ where
\bqa
{\rm E}[dW(t)] &=& 0 ,\\
dW(t)^{2} &=& dt. 
\eqa
Moreover, the stochastic post-measurement conditioned state is given by
\beq
\rho_{X_{w}}(t+dt) = \cu{1+{\cal D}[\hat c]dt + {\cal H}[\hat c]dW(t)}\rho(t).
\label{there}
\eeq
Here for arbitrary operators $\hat A$ and $\hat B$
\bqa
{\cal D}[\hat A]\hat B &\equiv& \hat A\hat B\hat A\dg - 
\half\{\hat A\dg \hat A,\hat B\} ,\\
{\cal H}[\hat A]\hat B  &\equiv& (\hat A-{\rm Tr}[\hat A\hat B])\hat 
B + {\rm H.c.}.
\eqa
Averaging over the measurement results would give
\beq
d\rho = dt{\cal D}[\hat{c}]\rho,
\eeq
which does not necessarily preserve $\langle\hat{X}\rangle_{\rho(t)}={\rm 
Tr}[\hat{X}\rho(t)]$ because 
$\hat{c}$ need not commute with $\hat{X}$.

For continuous measurements we wish to consider a sequence of 
measurements as described above. However, for the purposes of deriving 
weak values, we wish to assume we have just a single such weak measurement, 
at time $t$. (This unrealistic assumption will be removed in 
Sec.~III.) Say the system was prepared at time $0$ in state 
$\ket{\psi(0)}$ 
and evolved unitarily to state $\ket{\psi}$ at time $t$. Let us 
use $\rho(t)$ for $\ket{\psi}\bra{\psi}$. The 
infinitely weak (i.e. infinitesimally strong) measurement  is then 
performed \cite{fn3}. Then at time $T$ a final projective measurement is made, 
and we wish to keep only those values of $X$ obtained for which the 
final result corresponds to the state $\ket{\phi(T)}$. Let 
$\ket{\phi}$ be that state 
evolved unitarily backwards in time from $T$ to just after $t$. 
Dropping the time arguments, we have in 
the notation of Sec.~I, 
\bqa
\hspace{-5ex}{\phantom{\an{X_{w}}}}_{\phi}\!\an{X_{w}}_{\psi} &=& {\rm E}[X_{w}|\phi,\psi] = 
\sum_{x} x {\rm Pr}[X_{w}=x|\phi,\psi] \\
&=& \sum_{x} x {\rm Pr}[X_{w}=x,\phi|\psi] / {\rm Pr}[\phi|\psi]. 
\label{backto}
\eqa
Here we have used a sum rather than an integral merely for convenience.

Now because the measurement is infinitely weak, it does not affect 
the denominator $D={\rm Pr}[\phi|\psi]$, which equals $\st{\ip{\phi}{\psi}}^{2}$. The 
numerator meanwhile evaluates to
\bqa
N &=& \sum_{x} x {\rm Pr}[\phi|X_{w}=x,\psi]{\rm Pr}[X_{w}=x|\psi] \\
&=& \sum_{x} x \bra{\phi} \rho_{x}(t+dt) 
\ket{\phi}{\rm Pr}[X_{w}=x|\psi] \\
&=& {\rm E}[X_{w} \bra{\phi}\rho_{X_{w}}(t+dt)\ket{\phi} ].
\eqa
Now from the above formulae (\ref{here})--(\ref{there}) it is easy to 
see that this evaluates to
\beq
\bra{\phi} \cu{\rho(t)\langle\hat{X}\rangle_{\rho(t)} + {\cal H}[\hat c] \rho(t) }\ket{\phi}
 + O(dt).
\eeq
Expanding the superoperator ${\cal H}$ and ignoring infinitesimals gives 
\beq
\bra{\phi} \cu{\hat c \rho(t) + \rho(t)\hat c\dg }\ket{\phi}.
\eeq
Substituting this into \erf{backto}, and using 
$\rho(t)=\ket{\psi}\bra{\psi}$, gives
\bqa
\hspace{-5ex}{\phantom{\an{X_{w}}}}_{\phi}\!\an{X_{w}}_{\psi} &=& \frac{ \bra{\phi} 
\hat c \ket{\psi}\ip{\psi}{\phi} + {\rm H.c.}}
{\st{\ip{\phi}{\psi}}^{2}}\\ 
&=& 2{\rm Re}\frac{\bra{\phi} 
\hat c \ket{\psi}}{\ip{\phi}{\psi}} .\label{Swv}
\eqa
In the case of QND measurements where $\hat{c} = \hat{X}/2$, this 
reduces to the AAV formula (\ref{weakval}). Putting in the unitary time 
evolution explicitly, and setting $T=t+\tau$, 
\beq \label{withU}
\hspace{-5ex}{\phantom{\an{X_{w}}}}_{\phi}\!\an{X_{w}(t)}_{\psi} = 2{\rm Re}\frac{\bra{\phi(T)} \hat U(\tau)
\hat c \hat U(t)\ket{\psi(0)}}{\bra{\phi(T)}\hat U(T)\ket{\psi(0)}}.
\eeq

\section{Weak Values: the General Theory}

We now generalize the above theory by removing many of the assumptions.
First, we allow the initial system state $\rho(0)$ to be mixed. 
Second, we allow arbitrary Markovian evolution \cite{GarZol00} 
of $\rho$ between 
measurements:
\beq \label{me1}
\dot{\rho} = {\cal L}\rho \equiv -i[\hat{H},\rho]
+ \sum_{\mu} {\cal D}[\hat{c}_{\mu}].
\eeq
Here $\hat{H}$ is an Hermitian operator while 
$\cu{\hat{c}_{\mu}}_{\mu}$ is a set of arbitrary operators (which 
strictly should be bounded \cite{Lin76}). 
It would be natural for one of them to be the operator $\hat{c}$ 
introduced in Sec.~II. Then at any time between preparation and final 
measurement, the conditional evolution of the system would be 
\beq 
d\rho_{X_{w}(s)} = ds{\cal L}\rho   + dW(s){\cal H}[\hat{c}]\rho.
\eeq
Finally, we allow for the final measurement at time $T$ 
to be described by a 
positive operator $\hat{E}(T)$ which is not necessarily a projector 
$\ket{\phi(T)}\bra{\phi(T)}$. As long as $\hat{1}-\hat{E}(T) \geq 0$, 
$\hat{E}(T)$  
is an effect for some final measurement. 

Using $E$ and $\rho$ to denote the events of successful preparation 
and successful final measurement, the weak value at time $t$ 
we wish to calculate now is
\bqa
\hspace{-5ex}{\phantom{\an{X_{w}}}}_{E}\!\an{X_{w}}_{\rho} 
&=& {\rm E}[X_{w}|E,\rho] = 
\sum_{x} x {\rm Pr}[X_{w}=x|E,\rho] \\
&=& \sum_{x} x {\rm Pr}[X_{w}=x,E|\rho] / {\rm Pr}[E|\rho].
\eqa
As before, the denominator is unaffected by the weak measurement and 
is given by
\beq
D = {\rm Tr}[\hat{E}(T) e^{{\cal L}T} \rho(0)].
\eeq
The numerator can be manipulated similarly to above to yield
\bqa
N &=& {\rm E}\sq{ X_{w}(t) {\rm Tr}[\hat{E}(T) e^{{\cal L}\tau} 
\rho_{X_{w}}(t+dt) } \\
&=& {\rm Tr}[\hat{E}(T) e^{{\cal L}\tau} \cu{\hat{c}e^{{\cal 
L}t}\rho(0) + {\rm H.c.}}].
\eqa
Equation (\ref{withU}) thus generalizes to
\beq \label{genres}
\hspace{-5ex}{\phantom{\an{X_{w}}}}_{E}\!\an{X_{w}(t)}_{\rho} = 
2{\rm Re}\frac{{\rm Tr}[\hat{E}(T) e^{{\cal L}\tau} \cu{\hat{c}e^{{\cal 
L}t}\rho(0)}]}{{\rm Tr}[\hat{E}(T) e^{{\cal L}T} \rho(0)]}.
\eeq

The above result looks more like a normalized correlation function, 
as is familiar in quantum optics, than the unexpected AAV 
formula (\ref{weakval}). Indeed, it can be written in the \heix 
picture  \cite{fn4} as 
\beq \label{corfun}
\hspace{-5ex}{\phantom{\an{X_{w}(t)}}}_{E}\!\an{X_{w}}_{\rho} = 
\frac{\an{\hat{E}(T) \hat{c}(t) + 
\hat{c}\dg(t)\hat{E}(T)}}{\langle \hat{E}(T) \rangle},
\eeq
with initial conditions supplied by 
$\rho(0)$ and the bath state implied by the master equation 
(\ref{me1}). However, we can 
formulate \erf{genres} more like \erf{weakval} as follows. First we 
denote, in analogy to $\ket{\psi} = U(t)\ket{\psi(0)}$, 
\beq
\rho = e^{{\cal L}t} \rho(0).
\eeq
This is the usual forward time evolution by the master equation 
(\ref{me1}). Next we define a {\em retrodictive} \cite{BarPegJef00} 
positive operator  
$\hat{E}$ according to
\beq
{\rm Tr}[\hat{E}^{} \hat{A}] 
\equiv {\rm Tr}[\hat{E}(T) e^{{\cal L}\tau} \hat{A}],
\eeq
where $\hat{A}$ is an arbitrary operator. 

It is easy to verify that 
this $\hat{E}^{}$ is $\hat{E}(T)$ evolved forward by time $\tau$ 
according to the {\em 
retrodictive} master equation
\beq \label{retrome}
\frac{d \hat{E}^{}}{dt} = +i[\hat{H},\hat{E}^{}] + 
\sum_{\mu} \sq{\hat c_{\mu}\dg \hat{E}^{} 
\hat c_{\mu} - \half\{\hat c\dg_{\mu}\hat c_{\mu},\hat{E}^{}\} }
\eeq
Note that this differs slightly from the retrodictive master equation 
in Ref.~\cite{BarPegJefJed01}.   
This is because we do not require the norm of $\hat{E}^{}$
 to be preserved, as it appears in both the numerator and denominator of 
\erf{genres}. That \erf{retrome} does generate a positive operator is obvious 
from the solution as a ``Dyson-like'' sum constructed in the manner of 
Eq.~(7.24) of Ref.~\cite{Car93}. 
Using these definitions, we can write the postselected weak value as
\beq
\hspace{-5ex}{\phantom{\an{X_{w}}}}_{E}\!\an{X_{w}}_{\rho} = 2{\rm Re}
\frac{{\rm Tr}[\hat{E}^{} \hat{c} \rho]}{{\rm 
Tr}[\hat{E}^{}\rho]}.
\eeq
Thus retrodiction, quantum trajectories and weak values are all 
united in this expression.

\section{Weak Values in the Stony-Brook cavity QED experiment}

The preceding section showed that postselected 
weak values can be thought of as a correlation function 
(\ref{corfun}). In quantum optics weak measurements of the type 
described in Sec.~II are realized through homodyne detection 
\cite{Car93,WisMil93c}, where 
$\hat{c}$ is proportional to the lowering operator of the radiating 
system. The typical correlation functions thus found are of the form
(in the \heix picture)
\bqa
\an{:\hat X(t+\tau) \hat X(t):} &=& 
\left\langle \hat{c}(t+\tau)\hat{c}(t) + \hat{c}\dg(t+\tau) \hat{c}(t) 
\right. \nl{+} \left.
 \hat{c}\dg(t)\hat{c}(t+\tau) + \hat{c}\dg(t)\hat{c}\dg(t+\tau) 
 \right\rangle. \nn \\ && \label{firstorder}
\eqa
Usually they are evaluated at steady state (ss) where 
\beq
\rho(0) = \rho_{\rm ss} \propto {\rm nullspace}({\cal L}).
\eeq
Note that this is not of the form of \erf{corfun}, so that this 
correlation function cannot be thought of as a weak value in the AAV 
sense.

A notable exception to this typicality is the recent cavity QED experiment 
conducted at Stony-Brook \cite{FosOroCasCar00}. This involved a damped 
cavity, very 
weakly driven,  through which a beam of resonant atoms passed. The 
master equation for the system can be approximated as
\bqa
\dot{\rho} &=& \varepsilon[\hat a\dg-\hat a,\rho]+  2\kappa{\cal D}[\hat{a}]\rho 
\nl{+}\sum_{j}^{N} \cu{ g_{j}[\hat a\dg\hat\sigma_{j}\dg 
- \hat\sigma_{j}\dg\hat{a},\rho] +  
2\gamma_{\perp}{\cal D}[\hat\sigma_{j}]\rho}. \label{meSB}
\eqa
Here $\hat\sigma_{j} = \ket{g}_{j}\bra{e}$ is the lowering operator 
for the $j$th atom, and $\hat{a}$ is the annihilation operator for the 
field. The output from the cavity is split into two beams by an 
$85$--$15$ beam splitter, One beam
is detected by homodyne detection (with nett efficiency 
$\eta_{h} < 85\%$), and the other is detected using a photon 
counter (with nett efficiency $\eta_{c} < 15\%$). 

Choosing the slightly odd convention of measuring time in units of 
$1/(2\kappa\eta_{h})$, the homodyne measurement gives a current 
proportional to
\beq
X_{w}(t) = {\rm Tr}[\rho(t)\hat{X}] + \xi(t),
\eeq
where $\hat{X}=\hat{a}+\hat{a}\dg$. It 
conditions the system state $\rho$ by adding the term
\beq
dW(t){\cal H}[\hat{a}]\rho
\eeq
to $d\rho$. Thus the homodyne measurement gives a continuous weak 
measurement as in Sec.~II, with $\hat{c}=\hat{a}$. The photon counting 
also gives a continuous measurement that is weak in the sense that the 
average change in the conditioned system in time $\delta t$ is 
of order $\delta t$. However, unlike homodyne detection, sometimes the 
change is great. This can be seen from the set of effects 
$\{\hat{E}_{0},\hat{E}_{1}\}$ describing this measurement:  
\bqa
\hat{E}_{0} &=& 1 - \delta t \kappa \eta_{c} \hat{a}\dg\hat{a}  ,\\
\hat{E}_{1} &=& \delta t \kappa \eta_{c}\hat{a}\dg\hat{a}
\eqa
These give the probabilities for detecting zero or one photon 
over a short time $\delta t$. The latter result reveals a lot about 
the system, and its effect (and measurement operator) are not close to 
unity.

In the Stony-Brook experiment, the correlation between the homodyne 
current $X_{w}(t)$ and the photon count increment at time 
$T=t+\tau$ were measured 
in order to demonstrate an aspect of wave-particle duality. 
In the experiment $\tau$ ranged from negative to positive, and the 
measured correlation was proportional to (in the \heix picture)
\beq \label{corfun2}
h(t-T) = 
\frac{\an{:(\hat{a}\dg\hat{a})(T)\hat{X}(t):}_{\rm ss}}
{\an{\hat{a}\dg\hat{a}}_{\rm ss}}
\eeq
This is a ``fractional'' (3/2) order correlation function [in contrast to 
the first order function in \erf{firstorder}].
Here we are interested only in the case $\tau > 0$ in which case the 
correlation function can be written as \cite{fn2}
\beq
h(-\tau) = \frac{{\rm Tr}[\hat{E}_{1}(T) e^{{\cal L}\tau}(\hat{a}\rho_{\rm ss}+ 
\rho_{\rm ss}\hat{a}\dg)]}{{\rm Tr}[\hat{E}_{1}(T)\rho_{\rm ss}]}.
\eeq
Defining $\hat{E}_{1}$ to be $\hat{E}_{1}(T)$ retrodicted (as in 
Sec.~III) from time $T$ to time $t$, we can rewrite this correlation 
function as
\beq \label{SBwv1}
\hspace{-5ex}{\phantom{\an{X_{w}}}}_{E_{1}}\!\an{X_{w}}_{\rho_{\rm ss}} = 
2{\rm Re}\frac{{\rm Tr}
[\hat{E}_{1}\hat{a}\rho_{\rm ss}]}
{{\rm Tr}[\hat{E}_{1}\rho_{\rm ss}]} .
\eeq

It is now clear that the correlation function measured in 
Ref.~\cite{FosOroCasCar00} 
is a weak value, preselected by  the system being in its stationary state, 
and postselected on the the final measurement result ${E}_{1}$ (a 
photon detected at time $T$). Actually, since what was measured was a 
 correlation {\em function} (i.e. a function of $\tau$), this 
 experiment showed the {\em dynamics} of a weak value over time. 
 In this case, the spectrum of $\hat{X}$ 
is the real line, so there is no chance of observing a weak value 
outside of this range. Nevertheless, the weak values measured in the 
experiment were, for a range of times $\tau$, very far away from the 
stationary  average value of $\hat{X}$. This lead to the violation of 
various classical inequalities \cite{CarCasFosOro00,FosOroCasCar00}. 
In hindsight, the strangeness 
of the weak values in this experiment is not surprising, since 
the condition cited by Aharonov and Vaidman \cite{AhaVai90} is fulfilled. 
That is, 
the postselection is done on a rare event (the detection of one 
photon rather than zero photons).

There are a number of features of the above system that let us 
transform the generalized weak value (\ref{SBwv1}) into a form almost 
identical to the special form (\ref{Swv}) derived in Sec.~II. Because 
the driving is very weak ($\epsilon \ll \kappa,\gamma_{\perp}$), 
the ``jump'' terms in the master equation 
(\ref{meSB}) have a very small effect. This means that it is possible 
to approximate the Liouvillian as
\beq
{\cal L} = {\cal H}[-i\hat{H}_{\rm eff}],
\eeq
where
\bqa
-i\hat{H}_{\rm eff} &=& \varepsilon(\hat a\dg-\hat a) -  
\kappa\hat{a}\dg \hat{a} \nl{+} \sum_{j}^{N} \sq{
 g(\hat a\dg\hat\sigma_{j}\dg 
- \hat\sigma_{j}\dg\hat{a})  -  
\gamma_{\perp}\hat\sigma_{j}\dg\hat\sigma_{j}}.
\eqa
Ignoring normalization, this generates nonunitary evolution according 
to the operator
\beq
\hat N(t) = \exp(-i\hat{H}_{\rm eff}t).
\eeq
The stationary solution is, to a very good approximation, pure 
\cite{CarBreRic91}
\beq
\rho_{\rm ss} = \ket{\psi_{\rm ss}}\bra{\psi_{\rm ss}},
\eeq
where $\ket{\psi_{\rm ss}}$ is the eigenstate of $-i\hat{H}_{\rm eff}$ 
with eigenvalue having the largest real part. 

These results mean that we can approximate \erf{SBwv1} by
\beq
\hspace{-5ex}{\phantom{\an{X_{w}}}}_{E_{1}}\!\an{X_{w}}_{\psi_{\rm ss}} = 
2{\rm Re}\frac{\bra{\psi_{\rm ss}}
\hat{a}\dg\hat{a} \hat N(\tau) \hat{a}\ket{\psi_{\rm ss}}}
{\bra{\psi_{\rm ss}}\hat{a}\dg\hat{a}\hat N(\tau)\ket{\psi_{\rm ss}}}
\eeq
Again because of the weak excitation, $\ket{\psi_{\rm ss}}$ is, to a 
good approximation, given by
\beq
\ket{\psi_{\rm ss}} = \ket{0} + \alpha \ket{1} + \beta \ket{1'}.
\eeq
Here $\alpha$ and $\beta$ scale with $\epsilon$. The state $\ket{0}$ is the 
state with no excitations, $\ket{1}$ that with one photon but all 
atoms in the ground state, and $\ket{1'}$ a state with one collective 
atomic excitation but no photons. The same sort of expansion holds 
for $\hat{a}\ket{\psi_{\rm ss}}/\sqrt{\an{\hat{a}\dg\hat{a}}_{\rm ss}}$ 
(although the presence of a 
nonzero $\ket{1}$ term here indicates that $\ket{\psi_{\rm 
ss}}$ does have higher order terms). Furthermore, to leading order, 
$\hat N(t)$ preserves this expansion (apart from normalization). 
For these reasons, we can replace 
$\hat{a}\dg\hat{a}$, which is really $\hat{a}\dg \hat{a}\otimes 
\hat 1_{\rm atoms}$, by $\ket{1}\bra{1}$. That is, postselection on the 
detection of a photon is equivalent to postselection on projection into 
the state $\ket{1}$. The final result thus is
\beq \label{finres}
\hspace{-5ex}{\phantom{\an{X_{w}}}}_{E_{1}}\!\an{X_{w}}_{\psi_{\rm ss}} = 
2{\rm Re}\frac{\bra{1} \hat N(\tau) 
\hat{a}\ket{\psi_{\rm ss}}}
{\bra{1}\hat N(\tau)\ket{\psi_{\rm ss}}}. 
\eeq
This is completely analogous \erf{weakval}  
when we recognize $N\dg(\tau)\ket{1}$ as the unnormalized 
retrodicted final state. 

When $\ket{\psi_{\rm ss}}$ and $\hat N(\tau)$ are substituted into 
\erf{finres}, 
and only the terms of leading order in the driving $\epsilon$ are 
kept, the result is 
\beq \label{analcorfun}
\frac{\hspace{-5ex}{\phantom{\an{X_{w}}}}_{E_{1}}\!\an{X_{w}}_{\psi_{\rm 
ss}}}{\bra{\psi_{\rm ss}}\hat{X}\ket{\psi_{\rm ss}}}= 1 + \zeta 
e^{-\eta\tau }\sq{\cos\Omega \tau + 
\frac{\eta}{\Omega}\sin\Omega\tau},
\eeq
where $\eta = (\kappa+\gamma_{\perp})/2$ and 
where $\zeta$ and $\Omega$ are functions of $\kappa$, 
$\gamma_{\perp}$, and the coupling coefficients 
$\cu{g_{j}}_{j}$. This is
exactly the same as the analytical result in 
Ref.~\cite{FosOroCasCar00}, where this 
result was derived from considering the correlation function 
(\ref{corfun2}) with $T < t$. That is, with the photodetection 
preceding the homodyne measurement. In that case, the correlation 
function is
\beq
h(-\tau) = h(|\tau|) = \frac{{\rm Tr}[\hat{X} e^{{\cal L}|\tau|} 
\hat{a}\rho_{\rm 
ss}\hat{a}\dg]}{{\rm Tr} [\hat{a}\rho_{\rm 
ss}\hat{a}\dg]} .
\eeq
In Ref.~\cite{FosOroCasCar00} the symmetry of $h(\tau)$ was 
established by assuming Gaussian fluctuations.  
The analysis in this paper shows that the symmetry (to 
leading order) of $h(\tau)$ can be demonstrated explicitly by calculating 
both cases. The case of negative $\tau$ is found simply from 
the usual predictive quantum mechanics with wavefunction collapse 
\cite{CarBreRic91}, and the case of positive 
$\tau$ equally simply by retrodictive 
quantum mechanics with weak values, as here.

Finally, it is important to point out that the above form for the weak 
value as a function of time (\ref{analcorfun}) matches well with 
the experiment \cite{FosOroCasCar00}. This is so even though the 
experimental conditions are far from ideal, with fluctuations in the 
number of atoms, transverse motion of the atoms, imperfect beam 
alignment and so on.  This
shows that a weak value is not as fragile as one might have 
thought, and does give quantitative predictions for the behaviour of 
the system.

\section{Discussion}

In this paper I have combined quantum trajectory theory, and the 
concepts of weak values and retrodiction in order to show the relation 
between weak values and certain correlation functions as can be 
measured in  quantum optics. This required a generalization of the 
weak value theory of Aharonov, Albert and Vaidman. However, in a special case 
of considerable interest, the Stony-Brook wave-particle 
cavity QED experiment, I show that the correlation function reduces 
to a form as simple, and almost the same, as that originally derived 
by AAV. Moreover, this experiment measured not a weak value at a 
particular time, but the change in a weak value over time. That is, it 
showed the {\em dynamics} of a weak value. 

Weak values are of considerable intrinsic theoretical and 
philosophical interest. 
Moreover, the fact that they obey certain well-defined rules 
\cite{AhaVai90,Aha01} 
imply that they are also heuristic tools for predicting the results of 
experiments. Of course the results could be obtained by other, more 
laborious methods. But the power of having several fonts of intuition 
should not be undervalued. The analysis of Sec.~IV is a case in 
point.  This work should help to identify other instances 
in which the theory of weak values can be used in quantum optics and 
related areas. 
It may even suggest new experiments that provide further illustrations and 
applications for this thought-provoking area of quantum physics.


\end{multicols}

\begin{references}

\bibitem{AhaAlbVai88}
Y. Aharonov, D.Z. Albert, and L. Vaidman, 
Phys. Rev. Lett {\bf 60}, 1351 (1988).

\bibitem{AhaBerLeb64}
Y. Aharonov, O.G. Bergman, and J.L. Lebowitz, 
Phys. Rev. B {\bf 134}, 1410 (1964).

\bibitem{AhaVai90}
Y. Aharonov and L. Vaidman, 
Phys. Rev. A {\bf 41}, 11 (1990).

\bibitem{Har92}
L. Hardy, 
Phys. Rev. Lett. {\bf 68}, 2981 (1992).

\bibitem{Aha01}
Y. Aharonov {\em et al.}, quant-ph/0104062.

\bibitem{Mol01}
K. M\o lmer, quant-ph/0109042. 

\bibitem{fn1} 
Although, by deliberately constructing a weak 
measurement, the unusual predictions of AAV 
\cite{AhaAlbVai88} were   
experimentally verified not long after they were proposed, by
N.W.M. Ritchie, J.G. Story, and R.G. Hulet,
Phys. Rev. Lett. {\bf 66}, 1107 (1991).

\bibitem{Bel88}
V.P. Belavkin,
``Nondemolition measurement and nonlinear filtering of quantum
stochastic processes'', pp. 245-66 of
A. Blaqui\`ere (ed.),
{\em Lecture Notes in Control and Information
Sciences} {\bf 121} (Springer, Berlin, 1988).

\bibitem{BelSta92}
V.P. Belavkin and P. Staszewski,
Phys. Rev. A {\bf 45}, 1347 (1992).

\bibitem{Bar90}
A. Barchielli,
Quantum Opt. {\bf 2}, 423 (1990).

\bibitem{Bar93}
A. Barchielli,
Int. J. Theor. Phys. {\bf 32}, 2221 (1993).

\bibitem{DalCasMol92}
J. Dalibard, Y. Castin and K. M\o lmer,
Phys. Rev. Lett. {\bf 68}, 580 (1992).

\bibitem{GarParZol92}
C.W. Gardiner, A.S. Parkins, and P. Zoller,
Phys. Rev. A {\bf 46}, 4363 (1992).

\bibitem{Car93}
H.J. Carmichael,
{\em An Open Systems Approach to Quantum Optics}
(Springer-Verlag, Berlin, 1993).

\bibitem{WisMil93c} 
H.M. Wiseman and G.J. Milburn,
Phys. Rev. A {\bf 47}, 1652  (1993).

\bibitem{WalMil94}
D.F. Walls and G.J. Milburn,
{\em Quantum Optics}
(Springer, Berlin, 1994), and references therein.

\bibitem{FosOroCasCar00}
G.T. Foster, L.A. Orozco, H.M. Castro-Beltran, and H.J. Carmichael,
Phys. Rev. Lett. {\bf 85}, 3149 (2000).

\bibitem{Kra83}
K. Kraus,
{\em States, Effects, and Operations: Fundamental Notions of Quantum
Theory}
(Springer, Berlin, 1983).

\bibitem{BraKha92}
V.B. Braginsky and F.Y. Khalili,
{\em Quantum Measurement}
(Cambridge University Press, Cambridge, 1992).

\bibitem{GarZol00}
C.W. Gardiner and P. Zoller,
{\em Quantum Noise}
(Springer, Berlin, 2000).

\bibitem{WisDio01}
H.M. Wiseman and L. Di\'osi, 
J. Chem. Phys. {\bf 268}, 91 (2001). [quant-ph/0012016]

\bibitem{fn3}
Of course in a real experiment, such as Ref.~\cite{FosOroCasCar00}, 
the measurement is not infinitely weak, otherwise an infinitely large 
ensemble would be required to obtain a sensible average. The 
infinitesimal $dt$ is actually a finite but small $\delta t = 1/B$, 
where $B$ is the bandwidth of the apparatus (including the 
detector, amplifier, and often a deliberately introduced filter).

\bibitem{fn4}
Note that $\hat{E}(T)$ in the \heix picture is not 
necessarily the same as $\hat{E}(T)$ in the \schx picture. 

\bibitem{Lin76}
G. Lindblad,
Commun. Math. Phys. {\bf 48}, 199 (1976).

\bibitem{BarPegJef00}
S. M. Barnett, D. T. Pegg, and J. Jeffers, 
J. Mod. Opt. {\bf 47}, 1779 (2000).

\bibitem{BarPegJefJed01}
S.M. Barnett, D.T. Pegg,  J. Jeffers,  and O. Jedrkiewicz,
Phys. Rev. Lett. 86, 2455 (2001).

\bibitem{fn2}
Note that the $\tau$ I am using is the negative of the $\tau$ in 
Ref.~\cite{FosOroCasCar00}, which is why I write their correlation 
function as $h(-\tau)$. 

\bibitem{CarCasFosOro00}
H.J. Carmichael, H.M. Castro-Beltran, G.T. Foster, and L.A. Orozco,
Phys. Rev. Lett. {\bf 85}, 1855 (2000).

\bibitem{CarBreRic91}
H.J. Carmichael, R.J. Brecha, and P.R. Rice,
Opt. Comm. {\bf 82}, 73 (1991).



\end{references}
\end{document}